\documentclass[letterpaper, 10 pt, conference]{ieeeconf}  
\IEEEoverridecommandlockouts

\usepackage{amsmath,amssymb}
\usepackage{graphicx}
\usepackage{dsfont}

\newcommand{\cM}{\boldsymbol{{M}}}
\newcommand{\cMc}{\cM^{\mathsf{c}}}
\newcommand{\cMs}{\cM^{\mathsf{s}}}
\newcommand{\vx}{\boldsymbol{x}}
\newcommand{\vy}{\boldsymbol{y}}
\newcommand{\vc}{\boldsymbol{c}}
\newcommand{\F}{\mathds{F}}
\newcommand{\supp}{\mathsf{supp}}
\newtheorem{theorem}{Theorem}
\newtheorem{lemma}[theorem]{Lemma}

\newtheorem{prop}[theorem]{Proposition}

\newtheorem{defn}[theorem]{Definition}

\newtheorem{example}[theorem]{Example}

\begin{document}

\setlength{\pdfpageheight}{\paperheight}
\setlength{\pdfpagewidth}{\paperwidth}

\title{\LARGE \bf Compressed Sensing with Probabilistic Measurements: \\A Group
  Testing Solution}

\author{ Mahdi Cheraghchi, Ali Hormati, Amin Karbasi, Martin
  Vetterli\thanks{This work was supported by Swiss NSF grants
    200020-115983/1, 5005-67322 and 200020-103729.}\\
  EPFL, School of Computer and Communication Sciences\\
    1015 Lausanne, Switzerland \\
    \{mahdi.cheraghchi, ali.hormati, amin.karbasi,
    martin.vetterli\}@epfl.ch }



\maketitle
\thispagestyle{empty}
\pagestyle{empty}

\begin{abstract}
  Detection of defective members of large populations has been widely
  studied in the statistics community under the name ``group
  testing'', a problem which dates back to World War~II when it was
  suggested for syphilis screening. There, the main interest is to
  identify a small number of infected people among a large population
  using \emph{collective samples}. In viral epidemics, one way to
  acquire collective samples is by sending agents inside the
  population. While in classical group testing, it is assumed that the
  sampling procedure is fully known to the reconstruction algorithm,
  in this work we assume that the decoder possesses only
  \emph{partial} knowledge about the sampling process. This assumption
  is justified by observing the fact that in a viral sickness, there
  is a chance that an agent remains healthy despite having contact
  with an infected person. Therefore, the reconstruction method has to
  cope with two different types of uncertainty; namely, identification
  of the infected population and the partially unknown sampling
  procedure.

  In this work, by using a natural probabilistic model for ``viral
  infections'', we design non-adaptive sampling procedures that allow
  successful identification of the infected population with
  overwhelming probability $1-o(1)$. We propose both probabilistic and
  explicit design procedures that require a ``small'' number of agents
  to single out the infected individuals. More precisely, for a
  contamination probability $p$, the number of agents required by the
  probabilistic and explicit designs for identification of up to $k$
  infected members is bounded by $m = O(k^2 (\log n) / p^2)$ and
  $m = O(k^2 (\log^2 n) / p^2)$, respectively.  In both cases, a
  simple decoder is able to successfully identify the infected
  population in time $O(mn)$.
\end{abstract}

\section{Introduction}
Suppose that we have a large population in which only a small number
of people are infected by a certain viral disease (e.g., one may think
of a flu epidemic), and that we wish to identify the infected ones. By
testing each member of the population individually, we can expect the
cost of the testing procedure to be large.  If we could instead pool a
number of samples together and then test the pool collectively, the
number of tests required might be reduced.  This is the main
conceptual idea behind the classical \emph{group testing} problem
which was introduced by Dorfman \cite{ref:Dor43} and later found
applications in variety of areas. A few examples of such applications
include testing for defective items (e.g., defective light bulbs or
resistors) as a part of industrial quality assurance \cite{ref:SG59},
DNA sequencing \cite{ref:PL94} and DNA library screening in molecular
biology (see, e.g.,
\cite{ref:ND00,ref:STR03,ref:Mac99,ref:Mac99b,ref:CD08} and the
references therein), multiaccess communication \cite{ref:Wol85}, data
compression \cite{ref:HL00}, pattern matching \cite{ref:CEPR07},
streaming algorithms \cite{ref:CM05}, software testing
\cite{ref:BG02}, and compressed sensing \cite{ref:CM06}. See the books
by Du and Hwang \cite{ref:groupTesting,ref:DH06} for a detailed
account of the major developments in this area.

\begin{figure}[t]
  \centering
  \includegraphics[width=8.7cm]{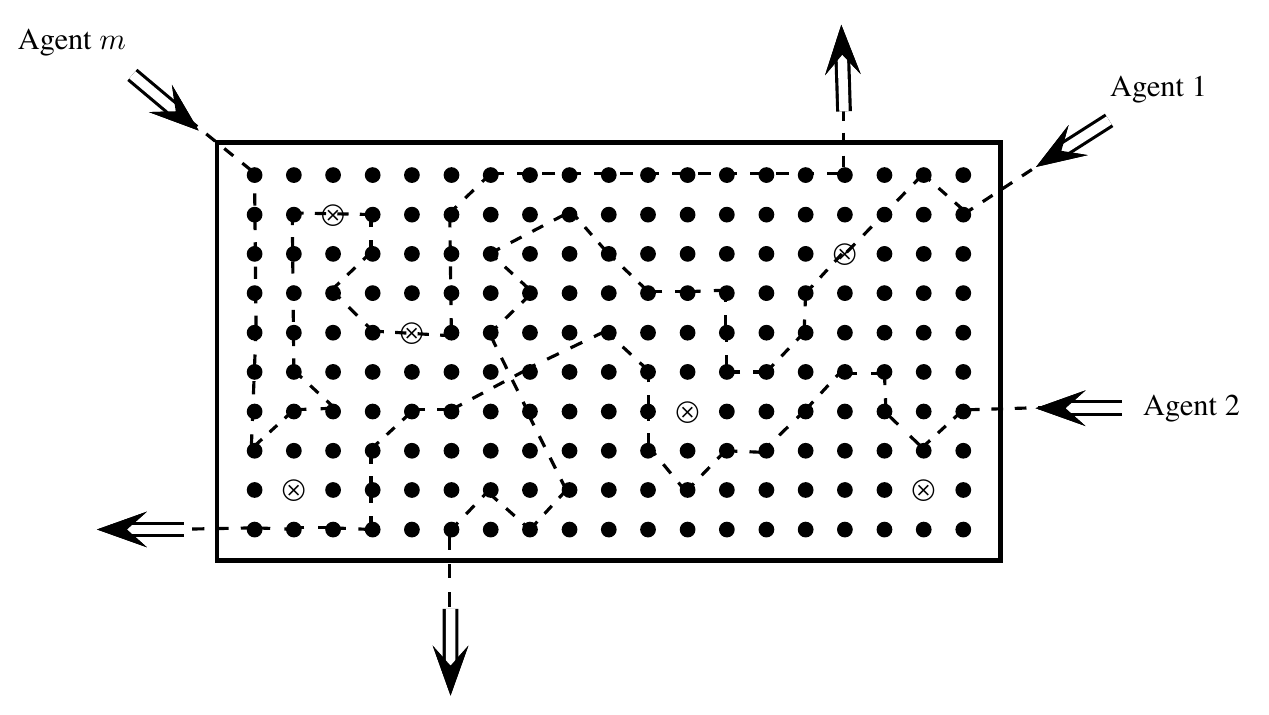}
  \caption{Collective sampling using agents. $\otimes$ symbols
    represent infected people among healthy people indicated by
    $\bullet$ symbols. The dashed lines show the individuals contacted by
    the agents. }\label{fig:agents}
\end{figure}

One way to acquire collective samples is by sending agents inside the
population whose task is to contact people (see
Fig.~\ref{fig:agents}). The agents can also be chosen as ATM machines,
cashiers in supermarkets, among other possibilities. Once an agent has
made contact with an ``infected'' person, there is a \textit{chance}
that he gets infected, too. By the end of the testing procedure, all
agents are gathered and tested for the disease. Here, we assume that
each agent has a \emph{log file} by which one can figure out with whom
he has made contact. One way to implement the log in practice is to
use identifiable devices (for instance, cell phones) that can exchange
unique identifiers when in range. This way, one can for instance ask
an agent to randomly meet a certain number of people in the population
and at the end learn which individuals have been met from the data
gathered by the device that is carried by the agent. Note that, even
if an agent contacts an infected person, he will not get infected with
certainty. Hence, it may well happen that an agent's result is
negative (meaning that he is not infected) despite a contact with some
infected person. We will assume that when an agent gets infected, the
resulting infection will not be contagious, i.e., an agent never
infects other people. Our ultimate goal is to identify the infected
persons with the use of a simple recovery algorithm, based on the test
results\footnote{In this work we focus on the exact reconstruction of
  the set of infected individuals in the worst case (i.e., regardless
  of the choice of this set).}. We remark that this model is
applicable in certain scenarios different from what we described as
well. For instance, in classical group testing, ``dilution'' of a
sample might make some of the items present in a pool ineffective. The
effect of dilution can be captured by the notion of contamination in
our model.

It is important to notice the difference between this setup and the
classical group testing where each contact with an infected person
will infect the agent with certainty. In other words, in the classical
group testing the decoder fully knows the sampling procedure, whereas
in our setup, it has only uncertain knowledge. Hence, in this scenario
the decoder has to cope simultaneously with two sources of
uncertainty, the unknown group of infected people and the partially
unknown (or stochastic) sampling procedure. 

The collective sampling can be done in adaptive or non-adaptive
fashions. In the former, samplings are carried out one at a time,
possibly depending the outcomes of the previous agents. However, in
the latter, the sampling strategy is specified and fixed before seeing
the the test outcome for any of the agents. In this paper we only
focus on non-adaptive sampling methods, which is more favorable for
applications.

The idea behind our setup is mathematically related to compressed
sensing \cite{CandesRT06, Donoho06}.  Nevertheless, they differ in a
significant way: In compressed sensing, the samples are gathered as
linear observations of a sparse real signal and typically tools such
as linear programming methods is applied for the reconstruction. To do
so, it is assumed that the decoder knows the measurement matrix a
priori. However, this is not the case in our setup. In other words,
using the language of compressed sensing, in our scenario the
measurement matrix might be ``noisy'' and is not precisely known to
the decoder. As it turns out, by using a sufficient number of agents
this issue can be resolved.

\section{ Problem Setting and Summary of the
  Results} \label{sec:setting}

To model the problem, we enumerate the individuals from $1$ to $n$ and
the agents from $1$ to $m$. Let the non-zero entries of $\vx := (x_1,
x_2, \dots, x_n) \in \F_2^n$ indicate the infected individuals within
the population. Moreover, we assume that $\vx$ is a $k$-sparse vector,
i.e., it has at most $k$ nonzero entries (corresponding to the
infected population).  We refer to the \emph{support set} of $\vx$ as
the the set which contains positions of the nonzero entries.

As typical in the literature of group testing and compressed sensing,
to model the non-adaptive samplings done by the agents, we introduce
an $m\times n$ boolean \textit{contact} matrix $\cMc$ where we set
$\cMc_{ij}$ to one if and only if the $i$th agent contacts the $j$th
person. As we see, the matrix $\cMc$ only shows which agents contact
which persons. In particular it does not indicate whether the agents
eventually get affected by the contact. Let us assume that at each
contact with a sick person an agent gets infected independently with
probability $p$ (a fixed parameter that we call the
\emph{contamination probability}).
Therefore, the real
\textit{sampling} matrix $\cMs$ can be thought of as a variation of
$\cMc$ in the following way:
\begin{itemize}
\item Each non-zero entry of $\cMc$ is flipped to $0$ independently
  with probability $1-p$;
\item The resulting matrix $\cMs$ is used just as in classical group
  testing to produce the \textit{outcome} vector $\vy \in \F_2^m$,
  \begin{equation}
    \vy=\cMs \cdot \vx,
  \end{equation}
  where the arithmetic is boolean (i.e., multiplication with the
  logical AND and addition with the logical OR).
\end{itemize}

The contact matrix $\cMc$, the outcome vector $\vy$, the number of
non-zero entries $k$, and the contamination probability $p$ are known
to the decoder, whereas the sampling matrix $\cMs$ (under which the
collective samples are taken) and the input vector $\vx$ are
unknown. The task of the decoder is to identify the $k$ non-zero
entries of $\vx$ based on the known parameters.

\begin{example}

  As a toy example, consider a population with $6$ members where only
  two of them (persons $3$ and $4$) are infected. We send three agents
  to the population, where the first one contacts persons $1,3,5$, the
  second one contacts persons $2,4,6$, and the third one contacts
  persons $2,3,5,6$. Therefore, the contact matrix and the input vector
  have the following form
  \begin{eqnarray*}
    \vx&=&(\begin{array}{c c c c c c}  0 & 0 & 1 & 1 & 0 & 0\end{array})^\top,\\
    \supp(\vx)&=&\{3,4\},\\
    \cMc&=&\left(\begin{array}{c c c c c c}
        1 & 0 & 1 & 0 & 1 & 0 \\
        0 & 1 & 0 & 1 & 0 & 1 \\
        0 & 1 & 1 & 0 & 1 & 1 \\
      \end{array}\right).
  \end{eqnarray*}
  Let us assume that only the second agent gets infected. This means
  that the outcome vector is
  \begin{equation*}
    \vy=(\begin{array}{c c c}  0 & 1 & 0 \end{array})^\top.
  \end{equation*}
  As we can observe, there are many possibilities for the sampling
  matrix, all of the following form:
  \begin{eqnarray*}
    \cMs&=&\left(\begin{array}{c c c c c c}
        ? & 0 & ? & 0 & ? & 0 \\
        0 & ? & 0 & ? & 0 & ? \\
        0 & ? & ? & 0 & ? & ? \\
      \end{array}\right),
  \end{eqnarray*}
  where the question marks are $0$ with probability $1-p$ and $1$ with
  probability $p$. It is the decoder's task to figure out which
  combinations make sense based on the outcome vector. For example,
  the following matrices and input vectors fit perfectly with $\vy$:
  \begin{eqnarray*}
    \left(\begin{array}{c}  0 \\ 1 \\0 \end{array}\right)&=&\left(\begin{array}{c c c c c c}
        1& 0 & 0 & 0 & 1 & 0 \\
        0 & 1 & 0 & 1 & 0 & 1 \\
        0 & 1 & 0 & 0 & 1 & 1 \\
      \end{array}\right) \left(\begin{array}{c}  0 \\ 0 \\ 1 \\ 1 \\ 0 \\ 0\end{array}\right),\\
    \left(\begin{array}{c}  0 \\ 1 \\0 \end{array}\right)&=&\left(\begin{array}{c c c c c c}
        1& 0 & 1 & 0 & 1 & 0 \\
        0 & 1 & 0 & 1 & 0 & 1 \\
        0 & 1 & 1 & 0 & 1 & 0 \\
      \end{array}\right) \left(\begin{array}{c}  0 \\ 0 \\ 0 \\ 1 \\ 0 \\ 1\end{array}\right).
  \end{eqnarray*}
\end{example}
\vspace{2mm}
More formally, the goal of our scenario is two-fold:
\begin{enumerate}
\item Designing the contact matrix $\cMc$ so that it allows unique
  reconstruction of \textit{any} sparse input $\vx$ from outcome $\vy$
  with overwhelming probability ($1-o(1)$) over the randomness of the
  sampling matrix $\cMs$.
\item Proposing a recovery algorithm with low computational
  complexity.
\end{enumerate}

In this work, we present a probabilistic and a deterministic approach
for designing contact matrices suitable for our problem setting along
with a simple decoding algorithm for reconstruction. Our approach is
to first introduce a rather different setting for the problem that
involves no randomness in the way the infection spreads out. Namely,
in the new setting an adversary can arbitrarily decide whether a
certain contact with an infected individual results in a contamination
or not, and the only restriction on the adversary is on the total
amount of contaminations being made. In this regard, the relationship
between the adversarial variation of the problem and the original
(stochastic) problem can be thought of akin to the one between the
combinatorial problem of designing block codes with large minimum
distances as opposed to designing codes for stochastic communication
channels.  The reason for introducing the adversarial problem is its
combinatorial nature that allows us to use standard tools and
techniques already developed in combinatorial group
testing. Fortunately it turns out that solving the adversarial
variation is sufficient for the original (stochastic) problem. We
discuss this relationship and an efficient reconstruction algorithm in
Section~\ref{sec:advers}.

Our next task is to design contact matrices suitable for the
adversarial (and thus, stochastic) problem. We extend two standard
techniques from group testing to our setting. Namely, we give a
probabilistic and an explicit construction of the contact matrix in
Sections \ref{sec:prob}~and~\ref{sec:expl}, respectively. The
probabilistic construction requires each agent to independently
contact any individual with a certain well-chosen probability and
ensures that the resulting data gathered at the end of the experiment
can be used for correct identification of the infected population with
overwhelming probability, provided that the number of agents is
sufficiently large.  Namely, for contamination probability $p$, we
require $O(k^2 (\log n) / p^2)$ agents, where $k$ is the estimate
on the size of the infected population. The explicit construction, on
the other hand, precisely determines which agent should contact which
individual, and guarantees correct identification with certainty in
the adversarial setting and with overwhelming probability (over the
randomness of the contaminations) in the stochastic setting.
This construction requires $O(k^2 (\log^2 n) / p^2)$ agents which is
inferior than what achieved by the probabilistic construction
by a factor $O(\log n)$.

We point out that, very recently, Atia and
Saligrama~\cite{ref:AtiaS09} developed an information theoretic
perspective applicable to a variety of group testing problems,
including a ``dilution model'' which is closely related to what we
consider in this work.  Contrary to our combinatorial approach, they
use information theoretic techniques to obtain bounds on the number of
required measurements. Their bounds are with respect to random
constructions and typical set decoding as the reconstruction
method. Specifically, in our terminology with contamination
probability $p$, they obtain an information theoretic upper bound of
$O(k^2 \log n / p^2)$ on the number measurements, which is comparable
to what we obtain in our probabilistic construction.


\vspace{2mm} {\it Remark:} As is customary in the standard group testing
literature, we think of the spartsity $k$ as a parameter that is
noticeably smaller than the population size $n$; for example, one may
take $k=O(n^{1/3})$.  Indeed, if $k$ becomes comparable to $n$, there
would be little point in using a group testing scheme and in practice,
for large $k$ it is generally more favorable to perform trivial tests
on the individuals.  Nevertheless it is easy to observe that our
probabilistic scheme can in general achieve $m=O(k^2 \log (n/k) / p^2)$,
but we ignore such refinements for the sake of clarity.

\section{Adversarial Setting} \label{sec:advers}

The problem described in Section~\ref{sec:setting} has a stochastic
nature, in that the sampling matrix is obtained from the contact
matrix through a random process. In this section we introduce an
adversarial variation of the problem that we find more convenient to
work with.

In the adversarial variation of the problem, the sampling matrix is
obtained from the contact matrix by flipping up to $e$ arbitrary
entries to $0$ on the support (i.e., the set of nonzero entries) of
each column of $\cMc$, for some \emph{error parameter} $e$.  The goal
is to be able to exactly identify the sparse vector despite the
perturbation of the contact matrix and regardless of the choice of the
altered entries.  Note that the classical group testing problem
corresponds to the special case $e=0$.  Thus the only difference
between the adversarial problem and the stochastic one is that in the
former problem the flipped entries of the contact matrix are chosen
arbitrarily (as long as there are not too many flips) while in the
latter they are chosen according to a specific random process.

It turns out that the combinatorial tool required for solving the
adversarial problem is precisely the notion of \emph{disjunct}
matrices that is well studied in the group testing literature. The
formal definition is as follows.

\begin{defn} \label{defn:disjunct} A boolean matrix $\cM$ with $n$
  columns $\cM_1, \ldots, \cM_n$ is called $(k,e)$-disjunct if, for
  every subset $S \subseteq [n]$ of the columns with $|S| \leq k$, and
  every $i \notin S$, we have
  \[
  \left|\supp(\cM_i) \setminus \left(\bigcup_{j \in S} \supp(\cM_j)
    \right)\right| > e,
  \]
  where $\supp(\cM_i)$ denotes the support of the column $\cM_i$.
\end{defn}

The following proposition shows a one-to-one correspondence between
contact matrices suitable for the adversarial problem and disjunct
matrices:

\begin{prop} \label{prop:disjunct} Let $\cM$ be a $(k, e)$-disjunct
  matrix. Then taking $\cM$ as the contact matrix solves the
  adversarial problem for $k$-sparse vectors with error parameter
  $e$. Conversely, any matrix that solves the adversarial problem must
  be $(k-1, e)$-disjunct.
\end{prop}

\begin{proof}
  Let $\cM$ be a $(k, e)$-disjunct matrix and consider $k$-sparse
  vectors $\vx, \vx'$ supported on different subsets $S, S' \subseteq
  [n]$.  Take an element $i \in S'$ which is not in $S$.  By
  Definition~\ref{defn:disjunct}, we know that the column $\cM_i$ has
  more than $e$ entries on its support that are not present in the
  support of any $\cM_j, j \in S$.  Therefore, even after $e$ bit
  flips in $\cM_i$, at least one entry in its support remains that is
  not present in the measurement outcome of $\vx'$, and this makes $\vx$
  and $\vx'$ distinguishable.

  For the reverse direction, suppose that $\cM$ is not $(k-1,
  e)$-disjunct and take any $i \in [n]$ and $S \subseteq [n]$ with
  $|S| \leq k-1$, $i \notin S$ which demonstrate a counterexample for
  $\cM$ being $(k-1, e)$-disjunct.  Consider $k$-sparse vectors $\vx$
  and $\vx'$ supported on $S$ and $S \cup \{i\}$, respectively. An
  adversary can flip up to $e$ bits on the support of $\cM_i$ from $1$
  to $0$, leave the rest of $\cM$ unchanged, and ensure that the
  measurement outcomes for $\vx$ and $\vx'$ coincide. Thus $\cM$ is not
  suitable for the adversarial problem.
\end{proof}

Of course, posing the adversarial problem is only interesting if it
helps in solving the original stochastic problem from which it
originates. Below we show that this is indeed the case; and in fact
the task of solving the stochastic problem reduces to that of the
adversarial problem; and thus after this point it suffices to focus on
the adversarial problem.

\begin{prop} \label{prop:advers} Suppose that $\cM$ is an $m \times n$
  contact matrix that solves the adversarial problem for $k$-sparse
  vectors with some error parameter $e$. Moreover, suppose that the
  weight of each column of $\cM$ is between $(1-\delta)qm$ and $qm$,
  for a parameter $q \in (0,1)$ and a constant $\delta \in (0,1)$, and that $e
  = (1-p)(1+\delta)qm$, for a constant $p \in (0, 1)$. Then $\cM$ can be
  used for the stochastic problem with contamination probability $p$, and
  achieves error probability at most
$n 2^{-\Omega(qm)}$,
  where probability is taken over the randomness of sampling (and the
  constant behind $\Omega(\cdot)$ depends on $p$ and $\delta$).
\end{prop}

\begin{proof}
  Take any column $\cM_i$ of $\cM$, and let $w_i$ be its weight.
  After the bit flips, we expect the weight of the column to reduce to
  $pw_i$. Moreover, by Chernoff bounds, the probability that (for
  ``small'' $\delta$) the amount of bit flips exceeds $(1-p)w_i(1+\delta)$
  is at most
\[  \begin{array}{l}
  \exp(-\delta^2 (1-p)w_i/4) \leq
\\ \qquad \qquad\exp(-\delta^2(1-\delta) (1-p)qm/4) =
  2^{-\Omega(qm)}.
  \end{array}\]
  Thus, by a union bound, the probability that the amount of bit flips
  at some column is not tolerable by $\cM$ is at most $n
  2^{-\Omega(qm)}$.
\end{proof}

\vspace{2mm} {\it Remark:} Note that, as we mentioned earlier, the
adversarial problem is stronger than classical group testing, and
thus, any lower bound on the number of measurements required for
classical group testing applies to our problem as well.  It is known
that any measurement matrix that avoids confusion in standard group
testing requires at least $\Omega(k^2 \log_k n)$ measurements
\cite{ref:DR83,ref:Rus94,ref:Fue96}.  Thus we must necessarily have $m
= \Omega(k^2 \log_k n)$ as well, and this upper bounds the error
probability given by Proposition~\ref{prop:advers} by at most
$n^{1-\Omega(qk^2/\log k)} = o(1)$.

%

\subsection{Decoding}
Suppose that the contact matrix $\cMc$ is $(k,
e)$-disjunct. Therefore, by Proposition~\ref{prop:disjunct} it can
combinatorially distinguish between $k$-sparse vectors in the
adversarial setting with error parameter $e$. In this work we consider
a very simple decoder that works as follows.

\vspace{2mm}
\noindent \textbf{Distance decoder:} For any column $\vc_i$ of the
contact matrix $\cMc$, the decoder verifies the following:
\begin{equation}\label{decoder}
  |\supp(\vc_i)\setminus\supp(\vy)|\leq e,
\end{equation}
where $\vy$ is the vector consisting of the measurement outcomes.
The coordinate $x_i$ is decided to be nonzero if and only if the
inequality holds.

\begin{lemma}
  The distance decoder correctly identifies the correct support of any
  $k$-sparse vector (with the above disjunctness assumption on $\cM$).
\end{lemma}

\begin{proof}
  Let $\vx$ be a $k$-sparse vector and $S := \supp(\vx)$, $|S| \leq
  k$, and $\cMc_S$ denote the corresponding set of columns in the
  sampling matrix.  Obviously all the columns in $\cMc_S$ satisfy
  \eqref{decoder} (as no column is perturbed in more than $e$
  positions) and thus the reconstruction includes the support of $\vx$
  (this is true regardless of the disjunctness property of $\cM$). Now
  let the vector $\hat{\vy}$ be the bitwise OR of the columns in
  $\cMc_S$ so that $\supp(\vy) \subseteq \supp(\hat{\vy})$, and assume
  that there is a column $\vc$ of $\cMc$ outside $S$ that satisfies
  \eqref{decoder}. Thus we will have
  $|\supp(\vc)\setminus\supp(\hat{\vy})|\leq e$, and this violates the
  assumption that $\cMc$ is $(k, e)$-disjunct.  Therefore, the
  distance decoder outputs the exact support of $\vx$.
\end{proof}


\section{Probabilistic Design} \label{sec:prob}

In light of Propositions \ref{prop:disjunct}~and~\ref{prop:advers}, we
know that in order to solve the stochastic problem with contamination
probability $p$ and sparsity $k$, it is sufficient to construct a $(k,
e)$-disjunct matrix for an appropriate choice of $e$. In this section,
we consider a probabilistic construction for $\cMc$, where each entry
of $\cMc$ is set to $1$ independently with probability $q :=
\alpha/k$, for a parameter $\alpha$ to be determined later, and $0$
with probability $1-q$. We will use standard arguments to show that,
if the number of measurements $m$ is sufficiently large, then the
resulting matrix $\cMc$ is suitable with all but a vanishing
probability.

Let $\delta > 0$ be an arbitrary (and small) constant.  Using Chernoff
bounds, we see that if $m \gg \log n$ (which will be the case), with
probability $1-o(1)$ no column of $\cMc$ will have weight greater than
$q(1+\delta)m$ or less than $q(1-\delta^2)m$. Thus in order to be able
to apply Proposition~\ref{prop:advers}, it suffices to set $e :=
(1-p)(1+3 \delta) qm$ as this value is larger than the error parameter
$(1-p)(1+\delta)^2qm$ required by the proposition.

\begin{lemma}
  For the above choices of the parameters $q$ and $e$, the probabilistic
  construction obtains a $(k, e)$-disjunct matrix with probability
  $1-o(1)$ using $m = O(k^2 (\log n) / p^2)$ measurements.
\end{lemma}

\begin{proof}
  Consider any set $S$ of $k$ columns of $\cMc$, and any column
  outside these, say the $i$th column where $i \notin S$. First we
  upper bound the probability of a \emph{failure} for this choice of
  $S$ and $i$, i.e., the probability that the number of the positions
  at the $i$th column corresponding to which all the columns in $S$
  have zeros is at most $e$.  Clearly if this event happens the $(k,
  e)$-disjunct property is violated.  On the other hand, if for no
  choice of $S$ and $i$ a failure happens the matrix is indeed $(k,
  e)$-disjunct.

  Now we compute the failure probability $p_f$ for a fixed $S$ and
  $i$.  A row is \emph{good} if at that row the $i$th column has a $1$
  but all the columns in $S$ have zeros. For a particular row, the
  probability that the row is good is $q(1-q)^k$. Then failure
  corresponds to the event that the number of good rows is at most
  $e$. The distribution on the number of good rows is binomial with
  mean $\mu = q(1-q)^k m$. By a Chernoff bound, the failure
  probability is at most
  \begin{eqnarray*}
    p_f &\leq& \exp( -(\mu-e)^2 / (2\mu)) \\
    &=& \exp(-mq ((1-q)^k - \\ && \qquad (1-p)(1+3\delta))^2/(2(1-q)^k)) \\
    &\leq& \exp(-mq (1/3^\alpha - (1-p)(1+3\delta))^2 / 2^{1-\alpha})
  \end{eqnarray*}
  where the last inequality is due to the fact that $(1-q)^k =
  (1-\alpha/k)^k$ is always between $1/3^\alpha$ and $1/2^\alpha$.
  Let $\gamma := (1/3^\alpha - (1-p)(1+3\delta))^2 / 2^{1-\alpha}$. Note
  that by choosing the parameters $\alpha$ and $\delta$ as
  sufficiently small constants, $\gamma$ can be made arbitrarily close
  to $p^2/2$.

  Now if we apply a union bound over all possible choices of $S$ and
  $i$, the probability of coming up with a bad choice of $\cMc$ would
  be at most $n \binom{n}{k} \exp(-mq \gamma)$. This probability
  vanishes so long as $m > k^2 \log (n/k) / (\alpha \gamma) = O(k^2
  (\log n) / p^2)$.
\end{proof}

Along with Propositions~\ref{prop:disjunct}~and~\ref{prop:advers}, the
result above immediately gives the following:

\begin{theorem}
  The probabilistic design for construction of an $m \times n$ contact
  matrix $\cMc$ achieves $m = O(k^2 (\log n) / p^2)$ measurements
  and error probability at most $n^{-\Omega(k/\log k)} = o(1)$ for the
  stochastic problem using distance decoder as the reconstruction
  method.
\end{theorem}

The probabilistic construction results in a rather sparse matrix,
namely, one with density $O(1/k)$ that decays with the sparsity
parameter $k$. Below we show that sparsity is necessary condition for the
construction to work:

\begin{lemma}
  Let $\cM$ be an $m \times n$ boolean random matrix, where $m = O(k^2
  \log n)$ for an integer $k > 0$, which is constructed by setting
  each entry independently to $1$ with probability $q$. Then either $q
  = O(\log k/k)$ or otherwise the probability that $\cM$ is
  $(k,e)$-disjunct (for any $e \geq 0$) approaches to zero as $n$
  grows.
\end{lemma}

\begin{proof}
  Suppose that $\cM$ is an $m \times n$ matrix that is
  $(k,e)$-disjunct.  Observe that, for any integer $t \in (0,k)$, if
  we remove any $t$ columns of $\cM$ and all the rows on the support
  of those columns, the matrix must remain $(k-t, e)$-disjunct. This
  is because any counterexample for the modified matrix being $(k-t,
  e)$-disjunct can be extended to a counterexample for $\cM$ being
  $(k,e)$-disjunct by adding the removed columns to its support.

  Now consider any $t$ columns of $\cM$, and denote by $m_0$ the
  number of rows of $\cM$ at which the entries corresponding to the
  chosen columns are all zeros. The expected value of $m_0$ is
  $(1-q)^t m$. Moreover, for every $\delta > 0$ we have
  \begin{equation} \label{eqn:chernoffDisj} \Pr[m_0 > (1+\delta)
    (1-q)^t m] \leq \exp( -\delta^2 (1-q)^t m/4 )
  \end{equation}
  by a Chernoff bound.

  Let $t_0$ be the largest integer for which $(1+\delta) (1-q)^{t_0} m
  \geq \log n$.  If $t_0 < k-1$, we let $t := 1+t_0$ above, and this
  makes the right hand side of \eqref{eqn:chernoffDisj} upper bounded
  by $o(1)$.  So with probability $1-o(1)$, the chosen $t$ columns of
  $\cM$ will keep $m_0$ at most $(1+\delta)(1-q)^t m$, and removing
  those columns and $m_0$ rows on their union leaves the matrix
  $(k-t_0-1, e)$-disjunct, which obviously requires at least $\log n$
  rows (as even a $(1, 0)$-disjunct matrix needs so many
  rows). Therefore, we must have
  \[
  (1+\delta)(1-q)^t m \geq \log n
  \]
  or otherwise (with overwhelming probability) $\cM$ will not be
  $(k,e)$-disjunct.  But the latter inequality is not satisfied by the
  assumption on $t_0$.  So if $t_0 < k-1$, little chance remains for
  $\cM$ to be $(k,e)$-disjunct.  Now consider the case $t_0 \geq
  k-1$. By a similar argument as above, we must have
  \[
  (1+\delta)(1-q)^k m \geq \log n
  \]
  or otherwise the matrix will not be $(k,e)$-disjunct with
  overwhelming probability.  The above inequality implies that we must
  have
  \[
  q \leq \frac{\log(m(1+\delta)/\log n)}{k},
  \]
  which, for $m = O(k^2 \log n)$ gives $q = O(\log k/k)$.
\end{proof}

\section{Explicit Design} \label{sec:expl} In the previous section we
showed how a random construction of the contact matrix achieves the
desired properties for the adversarial (and thus, stochastic) model
that we consider in this work. However, in principle an unfortunate
choice of the contact matrix might fail to be of use (for example, it
is possible though very unlikely that the contact matrix turns out to
be all zeros) and thus it is of interest to have an explicit and
deterministic construction of the contact matrix that is guaranteed to
work.

In this section, we demonstrate how a classical construction of
superimposed codes due to Kautz and Singleton \cite{ref:KS64} can be
extended to our setting by a careful choice of the parameters. This is
given by the following theorem.

\addtolength{\textheight}{-3.0cm} 

\begin{theorem}
  There is an explicit construction for an $m \times n$ contact matrix
  $\cMc$ that is guaranteed to be suitable for the stochastic problem
  with contamination probability $p$ and sparsity parameter $k$, and
  achieves $m = O(k^2 (\log^2 n) / p^2).$
\end{theorem}

\begin{proof}
  Let $m$ be an even power of a prime, and $n' := \sqrt{m}$. Consider
  a Reed-Solomon code of length $n'$ and dimension $k'$ over an
  alphabet of size $n'$. The contact matrix $\cMc$ is designed to have
  $n'^{k'}$ columns, one for each codeword.  Consider a mapping
  $\varphi\colon \F_{n'} \to \F_2^{n'}$ that maps each element of
  $\F_{n'}$ to a unique canonical basis vector of length $n'$; e.g.,
  $0 \mapsto (1,0,0,\ldots, 0)^\top$, $1 \mapsto (0,1, 0,\ldots,
  0)^\top$, etc.  The column corresponding to a codeword $\vc$ is set to
  the binary vector of length $m$ that is obtained by replacing each
  entry $c_i$ of $\vc$ by $\varphi(c_i)$, blowing up the length of $\vc$
  from $n'$ to $n'^2$.

  Note that the number of columns of $\cMc$ is $n := n'^{k'} =
  m^{k'/2}$, and each column has weight exactly $n' = m/n'$. Moreover,
  the support of any two distinct columns intersect at less than $k'$
  entries, because of the fact that the underlying Reed-Solomon code
  is an MDS code and has minimum distance $n'-k'+1$. Thus in order to
  ensure that $\cMc$ is $(k, e)$-disjunct, it suffices to have $ n' -
  kk' > e$ (so that no set of $k$ columns of $\cMc$ can cover too many
  entries of any column outside the set), or equivalently,
  \begin{equation} \label{eqn:MDS} \sqrt{m} - 2k (\log n / \log m) >
    e.
  \end{equation}
  By Proposition~\ref{prop:advers}, we need to set $e :=
  (1-p)(1+\delta)m/n'$ for an arbitrary constant $\delta > 0$. Thus in
  order to satisfy~\eqref{eqn:MDS}, it suffices to have $\sqrt{m}
  (1-(1-p)(1+\delta)) > 2k \log n$, which gives $m > 4k^2 \log^2 n /
  (1-(1-p)(1+\delta))^2$. As $\delta$ can be chosen arbitrarily small,
  the denominator can be made arbitrarily close to $p^2$ and thus we
  conclude that this construction achieves $m = O(k^2 \log^2 n / p^2)$
  measurements, which is essentially larger than the amount achieved
  by the probabilistic construction by a factor $O(\log n)$.
\end{proof}

Observe that, unlike the probabilistic construction of the previous
section, the explicit construction above guarantees a correct
reconstruction in the adversarial setting (where up to a $1-p$
fraction of the entries on the support of each column of the contact
matrix might be flipped to zero). Moreover, in the original stochastic
setting with contamination probability $p$, a single matrix given by the
explicit construction guarantees correct reconstruction with
overwhelming probability, where the probability is only over the
randomness of the testing procedure. This is in contrast with the
probabilistic construction where the failure probability is small, but
originates from two sources; namely, unfortunate outcome of the
testing procedure as well as unfortunate choice of the contact matrix
$\cMc$.

\bibliographystyle{IEEEtran} 


\end{document}